\documentclass[aip,graphicx,reprint]{revtex4-1}
\usepackage{graphicx}% Include figure files
\usepackage{dcolumn}% Align table columns on decimal point
\usepackage{bm}% bold math
\usepackage{upgreek} % upright greek letters in math mode
\usepackage{makecell} % multi line table entries
\usepackage[utf8]{inputenc}
\usepackage[T1]{fontenc}
\usepackage{mathptmx}
\usepackage{etoolbox}
\usepackage{xcolor}
\usepackage{amsmath}
\usepackage{natbib}
\usepackage{hyperref}

\makeatletter
\def\@email#1#2{%
 \endgroup
 \patchcmd{\titleblock@produce}
  {\frontmatter@RRAPformat}
  {\frontmatter@RRAPformat{\produce@RRAP{*#1\href{mailto:#2}{#2}}}\frontmatter@RRAPformat}
  {}{}
}%
\makeatother

\begin{document}

\preprint{AIP/123-QED}

\title[Influence of fluid rheology on multistability in the unstable flow of polymer solutions through pore constriction arrays]{Influence of fluid rheology on multistability in the unstable flow of polymer solutions through pore constriction arrays}
\author{Emily Y. Chen}
\author{Sujit S. Datta}
\email{ssdatta@princeton.edu}
\affiliation{ 
Department of Chemical and Biological Engineering, Princeton University, Princeton, NJ 08544, USA
}

\date{\today}
%%%%%%%%%%%%%%%%%%%%%%%%%%%%%%%%%%%%%%%%%%%%%%%%%%%
\begin{abstract}
Diverse chemical, energy, environmental, and industrial processes involve the flow of polymer solutions in porous media. The accumulation and dissipation of elastic stresses as the polymers are transported through the tortuous, confined pore space can lead to the development of an elastic flow instability above a threshold flow rate. This flow instability can generate complex flows with strong spatiotemporal fluctuations, despite the low Reynolds number ($\mathrm{Re} \ll 1$); for example, in 1D ordered arrays of pore constrictions, this unstable flow can be multistable, with distinct pores exhibiting distinct unstable flow states. Here, we examine how this multistability is influenced by fluid rheology. Through experiments using diverse polymer solutions having systematic variations in fluid shear-thinning or elasticity, in pore constriction arrays of varying geometries, we show that the onset of multistability can be described using a single dimensionless parameter. This parameter, the streamwise Deborah number, compares the stress relaxation time of the polymer solution to the time required for the fluid to be advected between pore constrictions. Our work thus helps to deepen understanding of the influence of fluid rheology on elastic instabilities, helping to establish guidelines for the rational design of polymeric fluids with desirable flow behaviors. 
\end{abstract}
%%%%%%%%%%%%%%%%%%%%%%%%%%%%%%%%%%%%%%%%%%%%%%%%%%%
\maketitle
%%%%%%%%%%%%%%%%%%%%%%%%%%%%%%%%%%%%%%%%%%%%%%%%%%%
\section{\label{sec:Intro}Introduction}\noindent A wide range of energy, environmental, industrial, and laboratory processes rely on the slow flow of solutions of large flexible polymers through porous media; examples include separations~\cite{ kozicki_filtration_1994,luo_high_1996,bourgeat_filtration_2003}, chemical production~\cite{petrie_instabilities_1976,denn_polymer_2008,turner_review_2014,elbadawi_polymeric_2018}, enhanced oil recovery~\cite{durst_flows_1981,seright_new_2010,sorbie_polymer-improved_2013,clarke_how_2016,pogaku_polymer_2018,mirzaie_yegane_fundamentals_2022}, groundwater remediation~\cite{roote_technology_1998,smith_compatibility_2008, huo_surfactant-enhanced_2020,hartmann_risk_2021}, and geothermal energy production~\cite{di_dato_impact_2022}. These processes require the spatiotemporal characteristics of the pore-scale flow to be predictable and controllable. However, the flow behavior typically depends on a complex interplay between the solution properties, imposed flow conditions, and porous medium geometry---all of which can vary greatly in practice---that is still poorly understood. Consequently, such processes often proceed by trial and error. Here, we take a step towards addressing this gap in knowledge by systematically studying how variations in polymer solution rheology influence flow in model porous media with precisely-defined geometries.

% Viscoelastic effects related to the onset of elastic flow instabilities~\cite{browne_homogenizing_2023, browne_harnessing_2023, clarke_how_2016, shakeri_effect_2021, mitchell_viscoelastic_2016, clarke_mechanism_2015, scholz_enhanced_2014} or increased extensional stresses~\cite{walker_enhanced_2014,kumar_stress_2023} have further been shown to enhance transport phenomena in confined environments compared to laminar flows of viscous, Newtonian fluids. Examples include improved mixing and subsequent speeding up of chemical reactions in porous media due to the onset of flow fluctuations from an elastic instability~\cite{browne_harnessing_2023} and improved heat transfer in a swirling flow and serpentine channel~\cite{traore_efficient_2015, whalley_enhancing_2015}. 

Such polymer solutions have two key rheological characteristics. First, they are often shear-thinning: the dynamic shear viscosity $\eta$ of a given solution decreases with increasing shear rate $\dot{\gamma}$, reflecting stretching and alignment of the constituent polymer chains under flow~\cite{macosko_rheology_1994,larson_structure_1999,rubinstein_polymer_2003,ryder_shear_2006}. Second, they are often highly elastic: the first normal stress difference $N_1$ of the solution is non-negligible and increases with shear rate in e.g., a cone-plate rheometer, reflecting the normal elastic stresses that arise as the polymer chains are stretched along the curved fluid streamlines~\cite{larson_instabilities_1992,mckinley_rheological_1996,pakdel_elastic_1996}. These elastic stresses can have dramatic consequences. A familiar example is the Weissenberg effect, in which the polymer solution ``climbs'' up a spinning rod inserted into it instead of being ejected away by inertia~\cite{weissenberg_continuum_1947,bird_dynamics_1987,more_rod-climbing_2023}. At sufficiently large flow speeds, these stresses accumulate faster than they can relax, causing the flow to become unstable---as exemplified in studies across a wide array of model geometries that feature curved streamlines~\cite{datta_perspectives_2022}. These studies have shown that such \emph{purely-elastic} instabilities --- termed such because they arise due to fluid elasticity, not inertia, at low Reynolds number $\textrm{Re}\ll 1$~\cite{pearson_instability_1976,larson_instabilities_1992,pakdel_elastic_1996,shaqfeh_purely_1996} --- can generate secondary flows~\cite{mckinley_rheological_1996,pakdel_elastic_1996,groisman_efficient_2001, ducloue_secondary_2019,shakeri_characterizing_2022}, eddies and vortices~\cite{batchelor_stress_1971,boger_viscoelastic_1987,koelling_instabilities_1991,byars_spiral_1994,mongruel_extensional_1995,khomami_stability_1997,arora_experimental_2002,mongruel_axisymmetric_2003,groisman_elastic_2000, rodd_role_2007,lanzaro_effects_2011,kenney_large_2013, gulati_flow_2015,shi_growth_2016,varshney_elastic_2017,qin_upstream_2019, hopkins_upstream_2022}, periodic flow fluctuations~\cite{sousa_purely-elastic_2018,hopkins_purely_2020}, dead zones~\cite{kawale_elastic_2017, kawale_polymer_2017,kawale_microfluidic_2019,ichikawa_viscoelastic_2022}, flow asymmetries~\cite{arratia_elastic_2006,galindo-rosales_optimized_2014,ribeiro_viscoelastic_2014,haward_elastic_2016,lanzaro_non-linear_2017,haward_steady_2018,davoodi_control_2019, qin_three-dimensional_2020, yokokoji_rheological_2023,kumar_stress_2023}, and even chaotic flows with a broad spectrum of spatial and temporal fluctuations~\cite{groisman_mechanism_1998,groisman_elastic_2000, groisman_efficient_2001,pan_nonlinear_2013,scholz_enhanced_2014,clarke_mechanism_2015,howe_flow_2015,lanzaro_quantitative_2015,traore_efficient_2015,whalley_enhancing_2015,clarke_how_2016,mitchell_viscoelastic_2016,qin_characterizing_2017,browne_elastic_2021, shakeri_effect_2021,carlson_volumetric_2022,  browne_homogenizing_2023,browne_harnessing_2023}, depending on the properties of the confining boundaries and imposed flow conditions.

To isolate the influence of fluid elasticity from shear-thinning, elastic instabilities are commonly studied using Boger fluids~\cite{boger_viscoelastic_1987,james_boger_2009} --- elastic but non-shear-thinning fluids composed of dilute amounts of the polymer dispersed in a highly-viscous solvent. However, the polymer solutions used in energy, environmental, industrial, and laboratory processes often differ from this idealized limit, exhibiting varying degrees of shear-thinning and elasticity~\cite{howe_flow_2015, clarke_how_2016, mirzaie_yegane_fundamentals_2022}. Studies in simplified geometries indicate that such variations in solution rheology can strongly influence how elastic instabilities manifest~\cite{larson_effect_1994, mckinley_rheological_1996,jagdale_fluid_2020,yokokoji_rheological_2023}, with some studies even suggesting that shear-thinning suppresses the onset of elastic instabilities altogether~\cite{larson_effect_1994,casanellas_stabilizing_2016,cagney_taylorcouette_2020,lacassagne_shear-thinning_2021}. However, decoupling the influence of shear-thinning and fluid elasticity is challenging in more geometrically-complex porous media in which the pore space geometry and thus, local flow conditions, are spatially highly heterogeneous. 

In previous work~\cite{browne_bistability_2020}, we used microfabricated one-dimensional (1D) ordered arrays of pore constrictions [schematized in Figure~\ref{experimental}(a)] to simplify this complexity. By directly visualizing the flow of an approximately Boger fluid through these arrays, we found that when the spacing between adjacent constrictions is sufficiently small, the unstable flow exhibits \emph{multistability}: it stochastically switches between distinct unstable flow states in the distinct pores. In the ``eddy-dominated'' state, large unstable eddies form in the corners of a given pore body in between adjacent constrictions; by contrast, in the ``eddy-free'' state, strongly fluctuating fluid pathlines fill the entire pore body and eddies do not form. Theoretical calculations, supported by the simulations of~\citeauthor{kumar_numerical_2021}, indicated that this unusual behavior arises from the competition between flow-induced polymer elongation, which promotes eddy formation~\cite{batchelor_stress_1971,boger_viscoelastic_1987,mongruel_extensional_1995,mongruel_axisymmetric_2003,rodd_role_2007}, and relaxation of polymers as they are advected between pore constrictions, causing elastic stresses to dissipate and enabling the eddy-free state to form. However, as in typical studies of elastic flow instabilities, this study used an elastic fluid that does not exhibit appreciable shear-thinning---despite the prevalence and importance of shear-thinning fluids in many real-world settings. Hence, we ask: How does shear-thinning influence the onset and features of this multistability?

Here, we address this question by experimentally studying the flow of polymer solutions with distinct rheological characteristics through 1D ordered arrays of pore constrictions. The solutions have systematic variations in either their degree of shear-thinning or fluid elasticity, enabling us to decouple the influence of these two rheological characteristics. Consistent with our prior work, we find that the fluid must be sufficiently elastic to become unstable and exhibit multistability. Moreover, we find that shear-thinning does not abrogate the onset of the elastic instability and the resulting development of multistability; however, it does influence the conditions at which multistability arises. In particular, for all polymer solutions tested, multistability arises when a characteristic stress relaxation time of the solution $\lambda$ approximately exceeds the characteristic time $t_{\mathrm{adv}}$ for fluid to be advected between pore constrictions; for non-shear-thinning solutions, $\lambda$ is given by the shear rate-independent longest stress relaxation time of the solution, whereas for shear-thinning solutions, $\lambda$ is instead rate-dependent. These results thus help expand understanding of flow multistability in porous media to a broader class of fluids, providing a way to use bulk rheology measurements to predict and control the pore-scale dynamics of unstable polymer solution flows. Not only does our work thereby deepen understanding of elastic instabilities, but it highlights a potentially useful way to harness such instabilities to alter momentum and mass transport in porous media~\cite{groisman_efficient_2001, burghelea_chaotic_2004, pathak_elastic_2004, scholz_enhanced_2014,traore_efficient_2015,whalley_enhancing_2015, kumar_stress_2023,browne_harnessing_2023}.
%%%%%%%%%%%%%%%%%%%%%%%%%%%%%%%%%%%%%%%%%%%%%%%%%%%
\section{\label{sec:Methods}Materials and Methods}

\subsection{Device fabrication}
\begin{figure}
\includegraphics{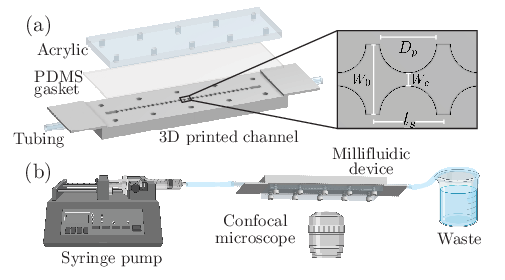}
\caption{\label{experimental} Schematic of the millifluidic device and experimental approach. (a) We fabricate millifluidic devices composed of a 3D-printed channel sealed by an overlying PDMS gasket and transparent acrylic top sheet, which allows for direct imaging of the flow within the channel. The inset shows the channel geometry: the square channel has width $W_0$ and height $H$, and pore constrictions of width $W_c$ are defined by opposing hemi-cylindrical posts of diameter $D_p$ spaced by a streamwise length $l_s$. (b) Each device is mounted on the stage of a confocal microscope used to directly image fluorescent tracers in the fluid, injected at a constant volumetric flow rate using a syringe pump. Figure created with \tt{biorender.com}.}
\end{figure}

\noindent We follow our previous work~\cite{browne_bistability_2020} in designing and fabricating the millifluidic devices used in the experiments. As shown in Fig.~\ref{experimental}(a), each device has a straight square channel $W_0=2$~mm wide and $H=2$~mm high with pore constrictions, evenly spaced by a distance $l_s$, defined by opposing hemi-cylindrical posts of diameter $D_p=1.6$~mm. To vary the extent to which elastic stresses can be retained between pore constrictions before relaxing, we test three different constriction-to-constriction spacings: $l_s=1W_0$, $1.3W_0$, and $1.6W_0$. The devices have either 30 ($l_s=W_0$) or 20 ($l_s=1.3W_0$ and $1.6W_0$) pore constrictions in total. For each device, we define the characteristic volume of a pore body as $V_{\mathrm{pore}}=HA_\mathrm{pore}$, where $A_\mathrm{pore}=l_sW_0-\pi D_p^2/4$.

We design each device using CAD software (Onshape) and 3D-print it using a proprietary clear polymeric resin made of methacrylate oligomers and photoinitiators (FLGPCL04) cured in a FormLabs Form 3 stereolithography 3D printer. Overlying the 3D-printed channel is a laser-cut clear acrylic top sheet fitted with screwholes (Epilog Mini 24). We sandwich a $\sim1$~mm thick sheet of polydimethylsiloxane (PDMS; Dow SYLGARD 184), made using a base-to-curing agent ratio of 8.5:1.5 by weight, between the 3D-printed channel and laser-cut acrylic top sheet to act as a gasket and ensure a watertight seal; each device is assembled by tightly screwing together the channel, PDMS, and acrylic layers. Finally, we glue flexible Tygon tubing (McMaster-Carr) into the inlets and outlets using a watertight two-part epoxy (JB MarineWeld).

\subsection{\label{subsec:formulation}Fluid formulations}\noindent We test eight different fluids, carefully formulated to have systematic variations in their degree of shear-thinning or fluid elasticity:
\begin{itemize}
\item Pure glycerol (Acros Organics)---a Newtonian, non-shear-thinning, non-elastic fluid that acts as a negative control.
\item 900 ppm xanthan gum (Sigma-Aldrich) dissolved in ultrapure (Milli-Q) water---a highly shear-thinning but not appreciably elastic fluid. This formulation is an entangled~\cite{ dobrynin_scaling_1995,howe_flow_2015,tran_relaxation_2023} semi-dilute polymer solution ($c/c^*\approx13$~\cite{wyatt_rheology_2009}), where $c^*$ is the overlap concentration at which adjacent polymer chains begin to interact with each other under quiescent conditions~\cite{rubinstein_polymer_2003}.
\item 300 ppm 30\% hydrolyzed $18$~MDa polyacrylamide (HPAM; Polysciences) dissolved in 89\% glycerol, 10\% ultrapure water, and 1\% NaCl (Sigma-Aldrich)---a not appreciably shear-thinning but highly elastic fluid. This formulation is a dilute polymer solution ($c/c^*\approx0.33$~\cite{browne_bistability_2020}). 
\item 300 ppm HPAM dissolved in 82.6\% glycerol, 10.4\% dimethyl sulfoxide (DMSO; Sigma-Aldrich), 6\% ultrapure water, and 1\% NaCl---another not appreciably shear-thinning but highly elastic fluid. This formulation is a dilute polymer solution ($c/c^*\approx0.5$~\cite{browne_elastic_2021}). 
\item 900 ppm HPAM dissolved in the same glycerol-DMSO-water-NaCl solvent---a moderately shear-thinning and highly elastic fluid. This formulation is an unentangled semi-dilute polymer solution ($c/c^*\approx1.5$~\cite{browne_elastic_2021}).
\item 4500 ppm HPAM dissolved in 89\% glycerol, 10\% ultrapure water, and 1\% NaCl---a highly shear-thinning and highly elastic fluid. This formulation is an unentangled semi-dilute polymer solution ($c/c^*\approx5$~\cite{browne_bistability_2020}). 
\item 1000 ppm HPAM dissolved in ultrapure water with 1\% NaCl---a moderately shear-thinning and moderately elastic fluid. This formulation is an unentangled semi-dilute polymer solution ($c/c^*\approx5$~\cite{liu_concentration_2009}). 
\item 3700 ppm HPAM dissolved in ultrapure water with 1\% NaCl---a highly shear-thinning and moderately elastic fluid. This formulation is an entangled semi-dilute entangled polymer solution ($c/c^*\approx19$~\cite{liu_concentration_2009}). 
\end{itemize}
The HPAM solutions have sufficient NaCl such that the ionic strength ($173$~mM) exceeds the charge concentration associated with the HPAM carboxylate groups for all concentrations tested; thus, the HPAM behaves as a flexible, neutral polymer due to excess salt screening in the formulations with NaCl~\cite{dobrynin_scaling_1995,howe_flow_2015}. As a shorthand, hereafter, we refer to aqueous solutions as Aq, glycerol-water solutions as Gl-Aq, and glycerol-water-DMSO solutions as Gl-Aq-DMSO.

\begin{figure}
\includegraphics{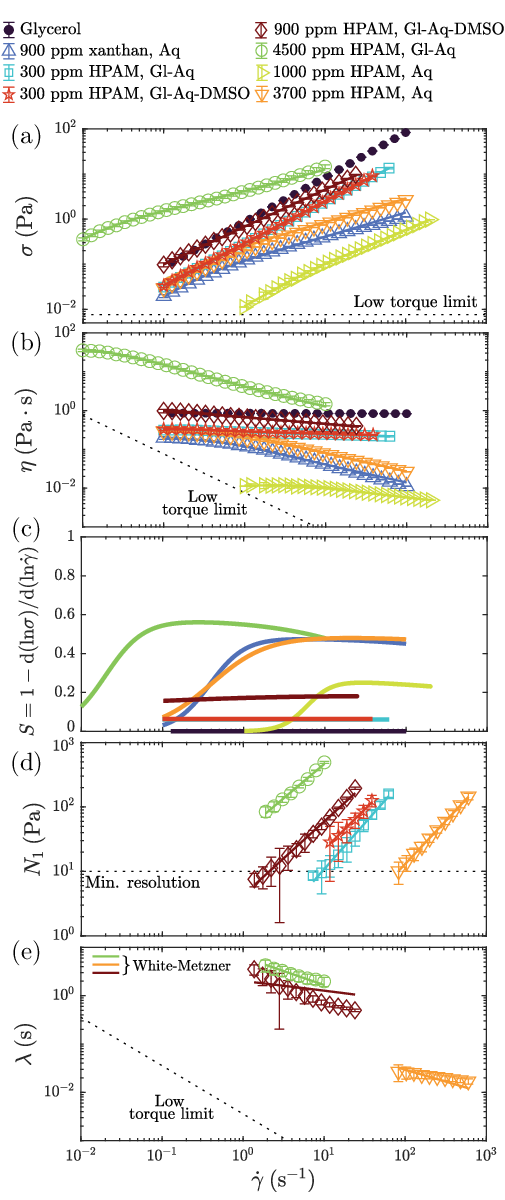}
\vspace{-10pt}
\caption{\label{rheology} Shear rheology of test solutions. (a) Shear stress $\sigma$ required to maintain an imposed shear rate $\dot{\gamma}$. Error bars show the standard deviation of 3 separate replicates of each solution. (b) Corresponding shear viscosity $\eta=\sigma/\dot{\gamma}$. Solid curves show Carreau-Yasuda or power-law fits, as described in the main text. (c) Corresponding shear-thinning parameter, $S$. (d) First normal stress difference, $N_1$. Solid curves show power-law fits. (e) Shear rate-dependent relaxation time, $\lambda(\dot{\gamma})$, determined as described in the main text. The sensitivity of the rheometer limits the resolution in measuring $N_1$, and correspondingly in determining $\lambda$ over all $\dot{\gamma}$ tested. Data for solutions with lower values of $N_1$ are not shown. Solid curves show White-Metzner fits.}
\end{figure}

To mix each polymer solution, we first dissolve the polymer in milliQ water in a conical tube on a rotor mixer for one day. We then dilute the solution with the remaining solvent components (glycerol, DMSO, and/or salt), and gently mix for at least 24 hours with a stir bar at 60 rpm to avoid mechanical degradation of the polymer. All polymer solutions are used within one month of mixing to avoid polymer degradation. We also seed each test fluid with 30~ppm of 1~$\upmu$m-diameter, carboxylate-modified polystyrene fluorescent tracer particles (Invitrogen) for flow visualization, as detailed further in \S\ref{subsec:flowvis}.

\begin{table*}
\caption{\label{tab:rheoparameters} 
Rheological parameters for the different test solutions, obtained using bulk shear rheology. Symbols are all defined in the main text. We estimate the zero-shear viscosity using the viscosity at the lowest measured shear rate, $\eta_0 \approx \eta|_{\dot{\gamma}=0.1 \: \mathrm{s^{-1}}}$ for the power law fluids.  We set $\eta_{\infty}=\eta_{s}$, the Newtonian solvent viscosity, for fitting as our measurements do not extend to sufficiently high rates to obtain a plateau. For the relaxation time, we use $\dot{\gamma} = 1 \: \mathrm{s^{-1}}$ as the lowest accessible shear rate within measurement limits to avoid noise from the instrument resolution at lower shear rates.}
\def\arraystretch{2.3}
\begin{ruledtabular}
\begin{tabular}{p{3cm}dddddddddddddd}
\multicolumn{1}{c}{\makecell{Solution}} & \multicolumn{1}{c}{$c/c^*$} & \multicolumn{1}{c}{\makecell{$\eta_0$ \\(Pa$\cdot$s)}} & \multicolumn{1}{c}{\makecell{$\eta_\infty$ \\(Pa$\cdot$s)}} & \multicolumn{1}{c}{\makecell{$K_\sigma$ \\ $(\mathrm{Pa\cdot s^{\emph{n}}})$}} & \multicolumn{1}{c}{$n$} & \multicolumn{1}{c}{\makecell{$\dot{\gamma}_c$ \\ $(\mathrm{s^{-1}})$}} & \multicolumn{1}{c}{$a$} & \multicolumn{1}{c}{$\beta$} & \multicolumn{1}{c}{\makecell{$\lambda_0$ \\ (s)}} & \multicolumn{1}{c}{\makecell{$K_{N_1}$ \\ $(\mathrm{Pa\cdot s^{\emph{n}_{\emph{N}_1}}})$}} & \multicolumn{1}{c}{$n_{N_1}$} & \multicolumn{1}{c}{\makecell{$\lambda_{0,\mathrm{WM}}$ \\(s)}} & \multicolumn{1}{c}{\makecell{$N_1 |_{10 \: \mathrm{s^{-1}}}$ \\ (Pa)}} & \multicolumn{1}{c}{$S_{\mathrm{max}}$}\\
\hline
\parbox{3cm}{Glycerol\\} & $---$ & 0.862 & $---$& $---$&$---$ & $---$& $---$&$---$ &$---$ & $---$&$---$ &$---$& $0$ & $0$ \\
\parbox{3cm}{900 ppm xanthan\\ Aq} & 13 & 0.196 & 0.001 & $---$ & 0.51 & 0.38 & 2 & 0.005 &$---$ &$---$ & $---$ & $---$ & $0$ & $0.47$  \\
\parbox{3cm}{300 ppm HPAM\\ Gl-Aq} & 0.3 & 0.334 & $---$ & 0.28 & 0.94 & $---$&$---$ & 0.66 & 2.1 & 0.41 & 1.4 & $---$ & $10.3$ & $0.06$ \\
\parbox{3cm}{300 ppm HPAM\\ Gl-Aq-DMSO} & 0.5 & 0.33 & $---$ & 0.29 & 0.94 & $---$ &$---$ & 0.19 & 2.2 & 1.3 & 1.2 & $---$ & $20.6$ & $0.06$ \\
\parbox{3cm}{900 ppm HPAM\\ Gl-Aq-DMSO} & 1.5 & 5.02 & 0.062 & $---$ & 0.73 & 0.006 & 0.15 & 0.01 & 10.6 & 4.3 & 1.1 & 16.3 & $54.1$ & $0.18$  \\
\parbox{3cm}{4500 ppm HPAM\\ Gl-Aq} & 5 & 38.8 & 0.22 & $---$ & 0.42 & 0.02 & 1.8 & 0.006 & 28 & 40.5 & 1.1 & 61.5 & $510$ & $0.56$  \\
\parbox{3cm}{1000 ppm HPAM\\ Aq} & 5 & 0.01 & 0.001 & $---$ & 0.71 & 6.1 & 2.8 & 0.09 & 0.2 & $---$ & $---$ & $---$ & $---$ & $0.25$  \\
\parbox{3cm}{3700 ppm HPAM\\ Aq} & 18.5 & 0.396 & 0.001 & $---$ & 0.51 & 0.4 & 1.3 & 0.003 & 2.0 & 0.03 & 1.3 & 0.44 & $0.60$ & $0.48$ \\
%300 ppm HPAM, Gl-Aq-DMSO & 0.5 & 0.33 & $---$ & 0.29 & 0.94 & $---$ &$---$ & 0.19 & 2.2 & 1.3 & 1.2 & $---$\\
%900 ppm HPAM, Gl-Aq-DMSO & 1.5 & 5.02 & 0.062 & $---$ & 0.73 & 0.006 & 0.15 & 0.12 & 10.6 & 4.3 & 1.1 & 16.3 \\
\end{tabular}
\end{ruledtabular}
\end{table*}

\subsection{\label{subsec:rheo}Bulk shear rheology} \noindent We characterize the shear rheology of each bulk solution using a stress-controlled Anton Paar MCR501 rheometer fitted with a truncated cone-plate geometry (CP50-2: $50$~mm diameter, $2^{\circ}$, $53~\upmu$m gap) and temperature-controlled at $25^{\circ}$C. In particular, we measure steady-state flow curves by ramping up, then ramping down, the imposed shear rate across the range $\dot{\gamma}=0.01-100 \: \mathrm{s^{-1}}$ and measure the shear stress $\sigma$ and first normal stress difference $N_1$. We do not observe substantial hysteresis with ramping direction in our measurements. The manufacturer-specified minimum torque is $T_{\mathrm{min}} = 0.05 \: \upmu$N$\cdot$m; we use $5\times$ this quoted minimum value to report the lower limit of resolvable stresses, viscosity, and relaxation time in our measurements. The lower limit of the measured $N_1$ is $\sim10$~Pa due to the normal force sensitivity.  For the dilute polymer concentrations, we use a lower shear rate limit of $\dot{\gamma}=0.1 \: \mathrm{s^{-1}}$ due to the minimum torque limitations. For solutions with large normal stresses, we only measure up to $\dot{\gamma} \approx 30 \: \mathrm{s^{-1}}$ to avoid elastic instabilities that develop in the cone-plate geometry.

\begin{figure}[h]
\includegraphics[width=0.5\textwidth]{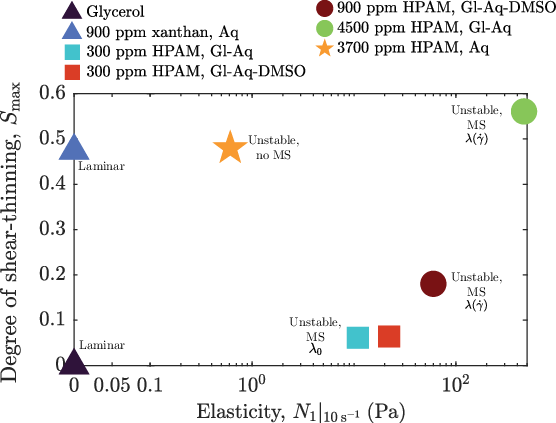}
% \vspace{-10pt}
\caption{\label{summary} Summary of different test fluids used, with systematically varying degrees of shear-thinning, quantified by the maximal measured shear-thinning parameter $S_{\mathrm{max}}$ from shear rheology, and elasticity, as quantified by the first normal stress difference measured at $\dot{\gamma}=10~\mathrm{s}^{-1}$, $N_1|_{10~\mathrm{s}^{-1}}$. `MS' refers to fluids that exhibit multistability in the experiments.}
\end{figure}

Our measurements are summarized in Fig.~\ref{rheology} and Table~\ref{tab:rheoparameters}. We first examine the shear-thinning nature of the different fluids. As classified in \S\ref{subsec:formulation}, the two 300 ppm HPAM solutions are not appreciably shear-thinning; as shown by the solid lines in Fig.~\ref{rheology}(a--b), the shear stress and viscosity vary as $\sigma(\dot{\gamma})= K_\sigma \dot{\gamma}^n$ and $\eta(\dot{\gamma})= K_\sigma \dot{\gamma}^{n-1}$, respectively, where $K_\sigma$ is known as the flow consistency index and the power law index $n=0.94\approx1$, indicating minimal shear-thinning. By contrast, the other polymer solutions show appreciable shear-thinning; as shown by the corresponding solid lines in Fig.~\ref{rheology}(a--b), the data are fit well by the Carreau-Yasuda model, $(\eta-\eta_0)/(\eta_{\infty}-\eta_0) = (1+(\dot{\gamma}/\dot{\gamma}_c)^a)^{(n-1)/a}$, where $\eta_0$ is the zero-shear viscosity, $\eta_{\infty}$ is the infinite-shear viscosity, $\dot{\gamma}_c$ is the critical shear rate for the onset of shear-thinning, and $a$ is a parameter that controls the transition to the shear-thinning regime. We further characterize this behavior in Fig.~\ref{rheology}(c) using the shear-thinning parameter $S \equiv 1- \frac{ \mathrm{d(ln}\sigma\mathrm{)}}{ \mathrm{d(ln}\dot{\gamma}\mathrm{)}}$, which is 1 minus the slope of the shear stress flow-curve on a log-log plot~\cite{haward_asymmetric_2020,yokokoji_rheological_2023}. For a non-shear-thinning fluid, the maximal measured $S_\mathrm{max} \approx 0$, while $0<S_\mathrm{max}<1$ for a shear-thinning fluid. As shown in Fig.~\ref{summary} and classified in \S\ref{subsec:formulation}, our solutions are either non-shear-thinning with  $S_\mathrm{max} \approx 0-0.05$ (glycerol, both 300~ppm HPAM solutions), moderately shear-thinning with $S_\mathrm{max} \approx 0.2-0.25$ (1000~ppm HPAM Aq and 900~ppm HPAM Gl-Aq-DMSO), or highly shear-thinning with $S_\mathrm{max} \approx 0.5$ (xanthan, 3700~ppm HPAM Aq, and 4500~ppm HPAM Gl-Aq).

Next, we examine the elasticity of the different fluids. Fig.~\ref{rheology}(d) shows the measured first normal stress difference with the solid lines indicating power law fits, $N_1(\dot{\gamma})=K_{N_1}\dot{\gamma}^{n_{N_1}}$, where $K_{N_1}$ is known as the consistency index and $n_{N_1}$ is the power law index for $N_1$. We use the value of $N_1$ measured at $\dot{\gamma}=10~\mathrm{s}^{-1}$, $N_1|_{10~\mathrm{s}^{-1}}$, as a simple way to characterize the extent of fluid elasticity. As shown in Fig.~\ref{summary} and classified in \S\ref{subsec:formulation}, glycerol and the xanthan solution are non-elastic with a non-measurable $N_1|_{10~\mathrm{s}^{-1}}\sim0$, while the 3700~ppm HPAM Aq solution is moderately elastic with $N_1|_{10~\mathrm{s}^{-1}}\sim1$~Pa, and the Gl-Aq and Gl-Aq-DMSO HPAM solutions are highly elastic with $N_1|_{10~\mathrm{s}^{-1}}\approx10-500$~Pa. The 1000~ppm HPAM Aq solution is also moderately elastic with $N_1|_{10~\mathrm{s}^{-1}}\lesssim1$~Pa; however, for this solution, the measured normal stress values are at the noise threshold of the rheometer, and we therefore omit them from Figs.~\ref{rheology}--\ref{summary} given the large measurement uncertainty.

We use shear stress relaxation measurements to characterize the longest relaxation time of the polymer solutions~\cite{liu_longest_2007,liu_concentration_2009}. In the cone-plate geometry, we impose a constant shear rate $\dot{\gamma} = 1 \: \mathrm{s^{-1}}$ operating in stress-controlled mode for $60$~s, stop shearing, and record the instantaneous shear stress response as it decays over time $t$. Fitting a single exponential decay as predicted by Maxwell relaxation model, $\sigma(t) \sim e^{-t/\lambda_0}$, to the linear portion of the stress-time curve in log-linear coordinates then yields an approximation to the longest relaxation time $\lambda_0$, whose values are given in Table~\ref{tab:rheoparameters}. This parameter describes the stress relaxation dynamics of non-shear-thinning dilute polymer solutions~\cite{boger_viscoelastic_1987,james_boger_2009}; however, for more concentrated solutions that exhibit shear-thinning, rheological properties exhibit strongly shear rate-dependent behavior~\cite{bird_dynamics_1987,casanellas_stabilizing_2016,shakeri_scaling_2022}. To characterize this rate dependence, we determine the relaxation time from the steady-state flow curves as $\lambda(\dot{\gamma})=\frac{N_1(\dot{\gamma})}{2\sigma(\dot{\gamma})\dot{\gamma}} \frac{\eta(\dot{\gamma})}{\eta(\dot{\gamma})-\eta_s}$ \cite{white_development_1963,bird_dynamics_1987,macosko_rheology_1994,casanellas_stabilizing_2016,qin_upstream_2019}; here, $\eta_s$ is the solvent viscosity, and we thereby define the parameter $\beta \equiv \eta_s/\eta_0$ to quantify the solvent contribution to the total solution viscosity. The corresponding data are shown in Fig.~\ref{rheology}(e). As shown by the solid curves, the data are described reasonably well by the White-Metzner model~\cite{white_development_1963}, $\lambda_{\mathrm{WM}}(\dot{\gamma})=\frac{\lambda_{0,\mathrm{WM}}}{(1+(\dot{\gamma}/\dot{\gamma}_c)^a)^{(1-n)/a}}$, where the parameters $\dot{\gamma}_c$, $a$, and $n$ are determined from the Carreau-Yasuda fits in Fig.~\ref{rheology}(b) and the characteristic relaxation time $\lambda_{0,\mathrm{WM}}$ is a fitting parameter. 

\subsection{Flow visualization}{\label{subsec:flowvis}}\noindent Prior to each flow experiment, we first flush the millifluidic device to be used with ultrapure (Milli-Q) water, ensuring that no air bubbles are retained in the channel. We then fill the device with pure glycerol, followed by the fluid to be tested at a constant low flow rate of $1.5 \: \mathrm{mL/hr}$ using a Harvard Apparatus PHD 2000 syringe pump for at least 3 hours to saturate the pore space. We then mount the millifluidic device on the stage of a Nikon A1R+ laser scanning confocal fluorescence microscope, positioning the set-up so that the syringe pump, device, and outlet waste jar are at the same height to avoid hydrostatic pressure differences [Fig.~\ref{experimental}(b)].

During each experiment, we progressively increase the inlet flow rate $Q$ from $0.5$ to $25$~mL/hr, injecting the test fluid for at least 90~min ($> 85 V_\mathrm{pore}$) at each flow rate before commencing imaging to ensure that the flow has reached a near-steady state. We report the results for each flow rate in terms of the shear rate at the constriction wall, $\dot{\gamma}_{w,c}$, defined in \S\ref{subsec:params} below. For each flow rate tested, we directly visualize the flow field in the millifluidic device using the confocal microscope, exciting the tracer particles seeded in the test fluid with a 488 nm laser and detecting their fluorescence emission using a 500-550 nm sensor. In particular, we use a $4\times$ objective lens to interrogate a two-dimensional $1583~\upmu\text{m}\times3167~\upmu\text{m}$ field of view at a pixel resolution of $\approx6~\upmu\text{m}$ and optical section thickness of $37~\upmu\text{m}$ at a fixed depth midway along the height of the channel. We acquire successive such images at a speed of 60~frames per second for two minutes per pore ($\geq 5 V_\mathrm{pore}$). These sequences of images generate the streakline videos shown in Supplementary Movies 1--2, for which we additionally time-average the intensity in each pixel over a running duration of 30 successive
frames to produce streaklines of the tracer particles. Furthermore, to represent the dynamic flow field in the static streakline images provided in the manuscript, we time-average the intensity in each pixel across successive images obtained over a total duration corresponding to $5V_{\mathrm{pore}}$. We use the streakline images to manually measure the combined sizes of any eddies that may arise in the upper and lower corners of a given pore body upstream of a constriction, $A_{\mathrm{eddy}}$.

\subsection{\label{subsec:params}Characteristic parameters describing flow}\noindent Given that the rheology of the test fluids is shear rate-dependent (Fig.~\ref{rheology}), we calculate the characteristic shear rates experienced by the fluids as they are transported through the millifluidic devices. The device channel width varies with streamwise position, $x$, as:
\begin{eqnarray*}
    W(x) = 
        \begin{cases}
            W_0-2\sqrt{(D_p/2)^2-(x-D_p/2)^2} & \text{for $0 \leq x < D_p$}\\
            W_0 & \text{for $D_p \leq x \leq l_s.$}\\
        \end{cases}
\end{eqnarray*}
The fluid interstitial velocity is then given by $U(x)=Q/HW(x)$, and we thereby define a position-dependent shear rate using the half-width of the channel as the characteristic length scale: $\dot{\gamma}(x)=2U(x)/W(x)=2Q/HW(x)^2$. The average shear rate in a pore is then given by $\langle \dot{\gamma} \rangle_x = \frac{1}{l_s} \int_{0}^{l_s} \dot{\gamma}(x) \mathrm{d}x.$ We also calculate the shear rate at the channel wall, $\dot{\gamma}_w(x)= \left(\frac{6Q}{WH^2}\right)\left(1+\frac{H}{W}\right)\left(\frac{2}{3}f^*\right)\left(\frac{b^*}{f^*}+\frac{a^*}{f^*}\frac{1}{n}\right)$, following~\citeauthor{harnett_heat_1989} and ~\citeauthor{son_determination_2007}; here, $a^*$, $b^*$, and $f^*$ are constants determined from numerical calculations that depend on the channel aspect ratio~\cite{harnett_heat_1989}. The wall shear rate takes on its maximal value, $\dot{\gamma}_{w,c}=\dot{\gamma}_w(W_c)$, at the pore constrictions with $a^* = 0.3475, \ b^* = 0.8444$, $f^* = 0.7946$, and $H/W_c = 5$.

The flow can then be described by four dimensionless parameters:
\begin{itemize}
    \item The Reynolds number comparing the strength of inertial to viscous stresses, $\mathrm{Re} = \rho U_c W_c/\eta_0$. Here, $\rho$ is the fluid density, $U_c=Q/HW_c$ is the average velocity in a pore constriction, and $W_c=0.4$ mm is the channel width at the constriction. Across all fluids and flow conditions tested, $\mathrm{Re}\lesssim10^{-1}$, indicating that inertial effects are negligible.

\item The Weissenberg number comparing the strength of elastic to viscous stresses, $\mathrm{Wi} = \frac{N_1(\dot{\gamma}_{w,c})}{2\sigma(\dot{\gamma}_{w,c})}$. Our experiments are characterized by $\mathrm{Wi}=0$ to $67$, indicating that elastic stresses can become sufficiently large to generate purely-elastic instabilities during flow. 

\item The Pakdel-McKinley number describing the loss of flow stability when sufficiently large elastic stresses propagate over sufficiently long timescales along curved streamlines~\cite{mckinley_rheological_1996,pakdel_elastic_1996}, $\mathrm{M} = \sqrt{2 \mathrm{Wi} \cdot\mathrm{De}_\mathcal{R}}$. Here, $\mathrm{De}_\mathcal{R}=\lambda/\left(\mathcal{R}/U_c\right)$, where $\mathcal{R} \approx (2/D_p + 32.5/W_0)^{-1}$ is the characteristic streamline radius of curvature~\cite{mckinley_rheological_1996}. Our experiments are characterized by $\mathrm{M}=0$ to $316$, indicating again that elastic stresses can become sufficiently large and persistent over time to generate purely-elastic instabilities during flow.

    \item The streamwise Deborah number comparing the fluid relaxation time to the characteristic time $t_{\mathrm{adv}}$ for fluid to be advected between pore constrictions, $\mathrm{De}=\lambda/t_{\mathrm{adv}}$. Here, $\lambda=\lambda_0$ or $\lambda_{\mathrm{WM}}(\dot{\gamma})$ for non-shear-thinning or shear-thinning fluids, respectively, and $t_{\mathrm{adv}}=V_c/Q$, where we define the characteristic volume of the straight channel extending between pores that is circumscribed by the cylindrical posts, $V_c=l_s W_c H$. As described further in \S\ref{subsec:deborah}, the central result of this paper is that $\mathrm{De}$ describes the onset of multistability across all fluids tested in this work.
\end{itemize}

%%%%%%%%%%%%%%%%%%%%%%%%%%%%%%%%%%%%%%%%%%%%%%%%%%%
\section{\label{sec:results}Results and Discussion}
\subsection{\label{subsec:300_MS}Multistability of a highly-elastic, non-shear-thinning fluid}
\begin{figure*}
\includegraphics{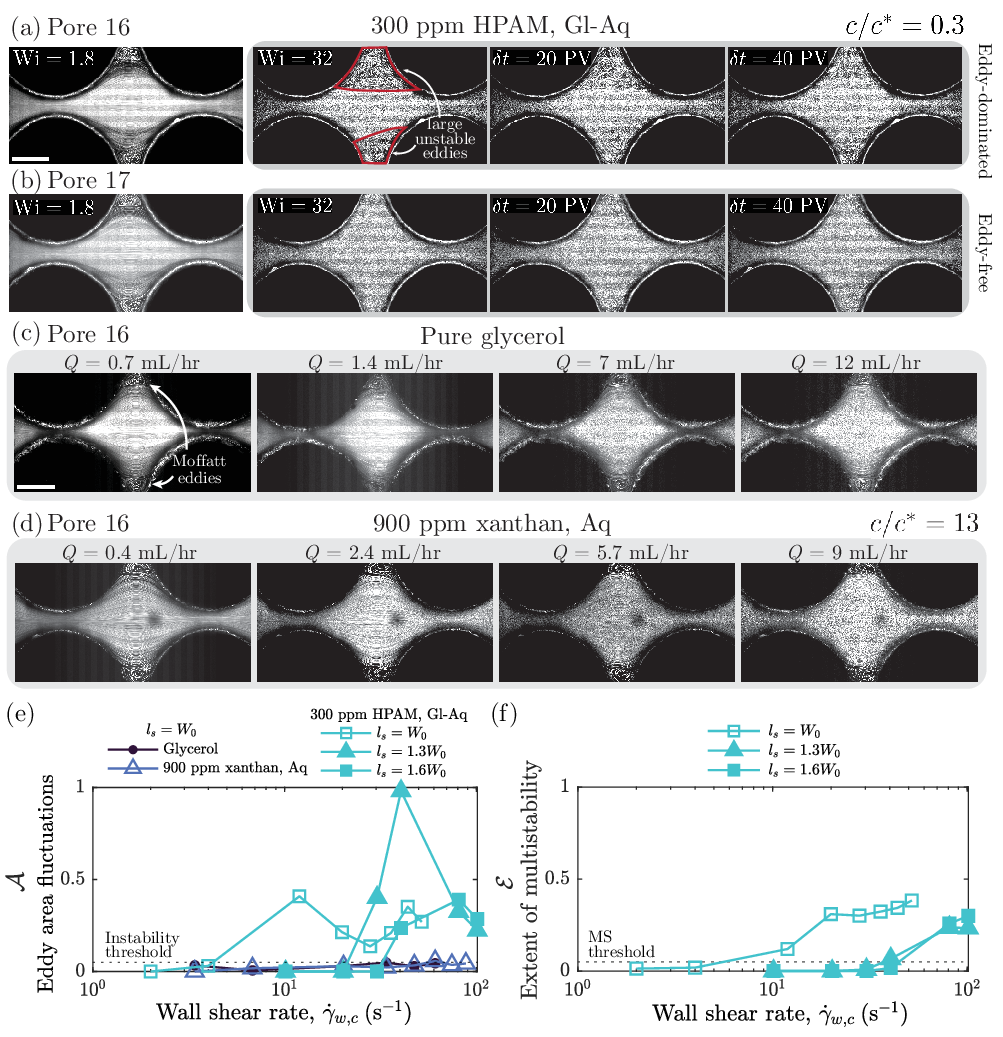}
\caption{\label{300hpam} Multistability of a highly-elastic, non-shear-thinning fluid (300 ppm HPAM, Gl-Aq). (a--b) Streakline images of the flow, averaged over a duration of 5 pore volumes (PVs), in pore bodies (a) 16 and (b) 17 at two different Weissenberg numbers, $\mathrm{Wi}=1.8$ (leftmost column) and $\mathrm{Wi}=32$ (right three columns showing three different time points separated by $\delta t=20$~PV). Scale bar represents 500~$\mathrm{\upmu m}$. At $\mathrm{Wi}=1.8$, the flow in both pores is laminar, with stable Moffatt eddies in the corners of the pore bodies. At $\mathrm{Wi}=32$, the flow is highly unstable, as shown by the crossing streaklines. During the imaging period, pore 16 is ``eddy-dominated'', with large fluctuating eddies located in the top and bottom corners of the pore body (outlined in red). At the same imposed flow rate, pore 17 is also unstable but exhibits  nearly ``eddy-free” behavior, with crossing streaklines filling more of the pore area. We term this finding that distinct pores exhibit distinct unstable flow states ``multistability''. (c-d) Streakline images of the flow of (c) glycerol and (d) xanthan solutions in pore body 16 over similar ranges of flow rates show laminar flow with constant Moffatt eddies in the corners of pore bodies, indicating that fluid elasticity is necessary for instability to occur. (e) We quantify the onset of unstable flow using the root-mean-square temporal fluctuations in the size of each eddy, $A'_{\mathrm{eddy,rms}}$, normalized by the mean value, $\langle A_{\mathrm{eddy}}\rangle_t$, aggregated for all imaged pores: $\mathcal{A}\equiv A'_{\mathrm{eddy,rms}}/\langle A_{\mathrm{eddy}}\rangle_t$. We characterize the flow as being unstable when $\mathcal{A}$ exceeds a threshold value $0.05$, determined from noise in eddy area measurements of glycerol and xanthan gum controls solutions that remain laminar. (f) We quantify the extent of multistability using the difference between the maximal and minimal observed eddy size, normalized by the size of a pore body: $\mathcal{E}\equiv\left(\mathrm{max}\{A_{\mathrm{eddy}}\}-\mathrm{min}\{A_{\mathrm{eddy}}\}\right)/A_{\mathrm{pore}}$. The onset of both unstable flow, and multistability, are pushed to higher shear rates as the pore constriction spacing increases.}
\end{figure*}

\noindent We first examine the flow of a highly-elastic, but non-shear-thinning, fluid (300 ppm HPAM, Gl-Aq---light blue square in Fig.~\ref{summary}) in a device with $l_s=W_0$. Streakline images of the flow in two neighboring pores near the middle of the device are shown in Figs.~\ref{300hpam}(a--b). As exemplified by the leftmost panels, at low flow rates, the flow is laminar throughout the entire pore space; it remains steady over time, with small Moffatt eddies in the corners upstream of each constriction. Above a threshold flow rate, however, the flow becomes unstable: the flow velocities fluctuate both spatially and temporally. An example is shown by the crossing streaklines in the righthand panels of Figs.~\ref{300hpam}(a--b), which represent the flow at three different times. We quantify the onset of this elastic instability by measuring the root-mean-square temporal fluctuations in the size $A_{\mathrm{eddy}}$ of each eddy, $A'_{\mathrm{eddy,rms}}$, where the prime indicates fluctuations from the mean value, $\langle A_{\mathrm{eddy}}\rangle_t$. The open square symbols in Fig.~\ref{300hpam}(e) show these measurements aggregated across multiple pore bodies of the medium. Consistent with the flow images shown in Figs.~\ref{300hpam}(a--b), $\mathcal{A}\equiv A'_{\mathrm{eddy,rms}}/\langle A_{\mathrm{eddy}}\rangle_t$ increases above the noise threshold at a constriction wall shear rate $\dot{\gamma}_{w,c} \sim4-10 \: \mathrm{s^{-1}}$, corresponding to $\mathrm{Wi}\sim2-6$ and $\mathrm{M}\sim6-19$. By contrast, Newtonian glycerol or a shear-thinning but non-elastic xanthan solution remain laminar at the same flow rates [Figs.~\ref{300hpam}(c--d)], confirming that the fluid must be sufficiently elastic to become unstable~\cite{pearson_instability_1976,larson_instabilities_1992,pakdel_elastic_1996,shaqfeh_purely_1996,datta_perspectives_2022}.

Even though all the pore constrictions in the array are fabricated to be geometrically identical, the features of the unstable flow differ from pore to pore. In particular, consistent with our previous findings~\cite{browne_bistability_2020}, individual pore bodies exhibit one of two distinct unstable flow states, each of which persists over long durations---a phenomenon we term \emph{multistability}---as exemplified by the righthand panels in Figs.~\ref{300hpam}(a--b) and in Supplementary Movie 1. Pore body 16 [top row] is ``eddy-dominated'' during the imaging period: large, fluctuating eddies [red outlines] form and persist in the corners between pore constrictions. Pore body 17 [bottom row] is instead ``eddy-free'': the fluctuating fluid pathlines fill most of the pore space and eddies do not persist in all corners between constrictions. We quantify this behavior by measuring the difference between the maximal and minimal observed eddy size, normalized by the size of a pore body: $\mathcal{E}\equiv\left(\mathrm{max}\{A_{\mathrm{eddy}}\}-\mathrm{min}\{A_{\mathrm{eddy}}\}\right)/A_{\mathrm{pore}}$, which characterizes the extent of multistability. As shown by the open square symbols in Fig.~\ref{300hpam}(f), multistability arises concomitant with the onset of unstable flow at a constriction wall shear rate $\dot{\gamma}_{w,c} \sim4-10 \: \mathrm{s^{-1}}$, corresponding to $\mathrm{Wi}\sim2-6$ and $\mathrm{M}\sim6-19$. 

\begin{figure*}
\includegraphics{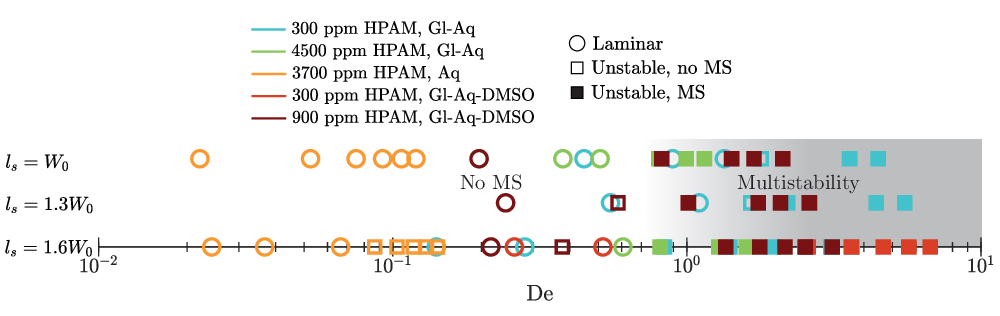}
\caption{\label{1D_state} Streamwise Deborah number parameterizes the onset of multistability. The data show all our measurements across different flow rates, test fluids of varying rheological properties, and medium geometries of varying pore constriction spacings. The Deborah number is defined as $\mathrm{De}=\lambda/t_{\mathrm{adv}}$, where $\lambda$ is a rate-dependent relaxation time and $t_{\mathrm{adv}}$ is the characteristic time for fluid to be advected between pore constrictions, as described in the main text. Open circles indicate experiments where the flow is laminar; open squares indicate experiments where the flow is unstable, but does not exhibit multistability; filled squares indicates experiments where the flow is both unstable and multistable. Across all experiments, the transition to multistability occurs at $\mathrm{De}\sim1$.}
\end{figure*}

Simulations~\cite{kumar_numerical_2021} indicate that this unusual behavior arises from the competition between flow-induced polymer elongation, which promotes eddy formation~\cite{batchelor_stress_1971,boger_viscoelastic_1987,mongruel_extensional_1995,mongruel_axisymmetric_2003,rodd_role_2007}, and relaxation of polymers as they are advected between pore constrictions, causing elastic stresses to dissipate and enabling the eddy-free state to form. To test this idea, we increase $l_s$, providing more time for elastic stresses to relax as fluid is advected between constrictions. In this case, we expect that the onset of multistability is suppressed and shifted to higher shear rates. Repeating our experiments for $l_s=1.3W_0$ and $1.6W_0$ confirms this expectation. The elastic instability arises at $\dot{\gamma}_{w,c} \sim20-30 \: \mathrm{s^{-1}}$ ($\mathrm{Wi}\sim11-17$, $\mathrm{M}\sim17-26$) and $\dot{\gamma}_{w,c} \sim30-40 \: \mathrm{s^{-1}}$ ($\mathrm{Wi}\sim17-24$, $\mathrm{M}\sim26-35$) for $l_s=1.3W_0$ and $1.6W_0$, respectively, as shown by the filled triangles and squares in Fig.~\ref{300hpam}(c). Correspondingly, multistability arises only above $\dot{\gamma}_{w,c} \sim30-40 \: \mathrm{s^{-1}}$ ($\mathrm{Wi}\sim55-73$, $\mathrm{M}17-24$) and $\dot{\gamma}_{w,c} \sim40-80 \: \mathrm{s^{-1}}$ ($\mathrm{Wi}\sim73-140$, $\mathrm{M}\sim24-52$) for $l_s=1.3W_0$ and $1.6W_0$, respectively, as shown by the filled triangles and squares in Fig.~\ref{300hpam}(d). Furthermore, repeating these experiments for another highly-elastic but non-shear-thinning fluid, but with a different solvent (300 ppm HPAM, Gl-Aq-DMSO---red square in Fig.~\ref{summary}), yields similar results [summarized in Fig.~\ref{1D_state} for brevity]---indicating that our findings are more general.

\subsection{\label{subsec:rheo_MS}Influence of shear-thinning on multistability}\noindent To investigate how fluid shear-thinning may influence the onset and features of this multistability, we next repeat the same experiments, but with different elastic fluids of systematically-varying degrees of shear-thinning. 

\begin{figure*}
\includegraphics{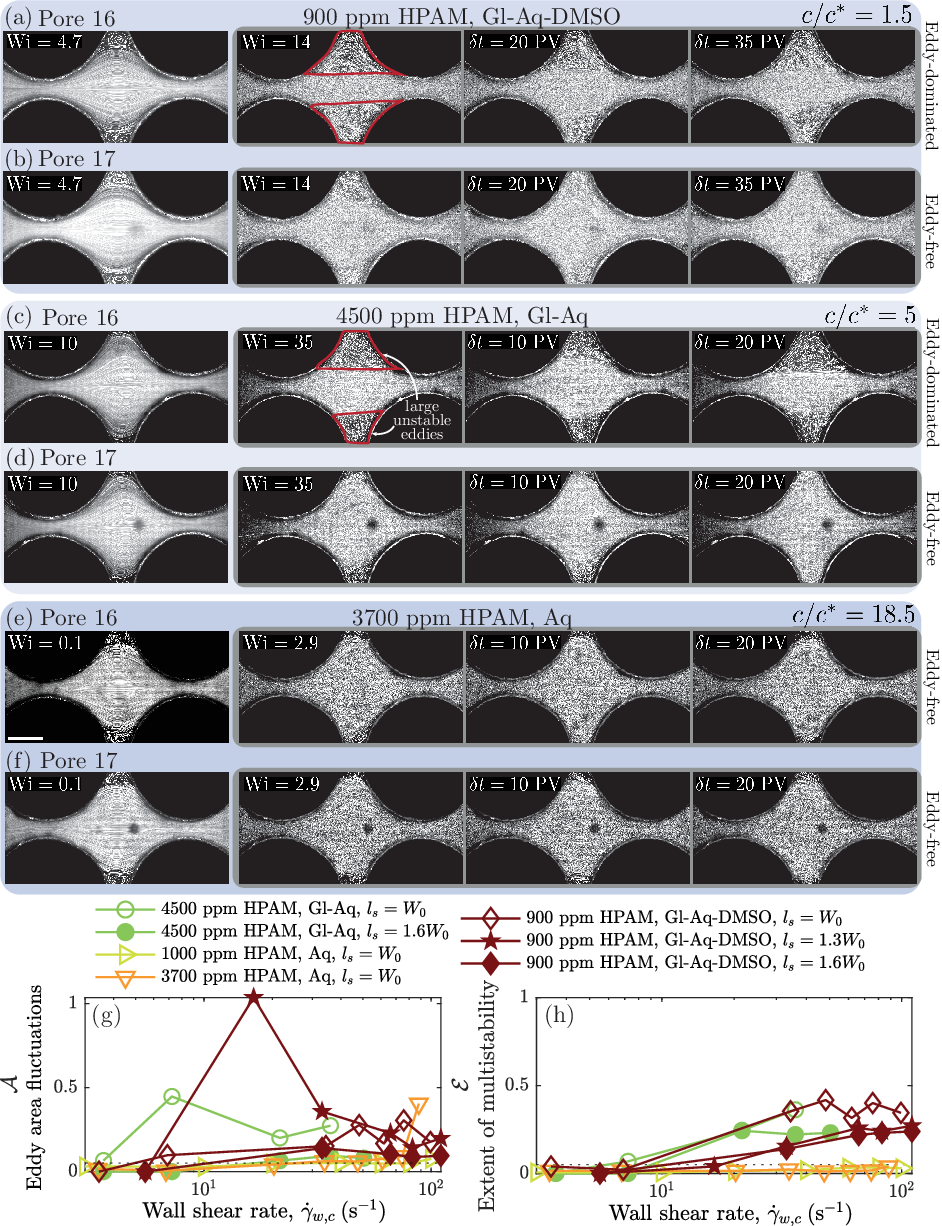}
\caption{\label{streakline_3700_hq900_4500}Multistability of fluids with different rheological properties. (a--f) Streakline images of the flow, averaged over 5 PVs, in pore bodies 16 and 17 at two different Weissernberg numbers, for (a--b) moderately shear-thinning and highly-elastic 900 ppm HPAM, Gl-Aq-DMSO, (c--d) highly shear-thinning and elastic 4500 ppm HPAM, Gl-Aq, and (e--f) highly shear-thinning and moderately-elastic 3700 ppm HPAM, Aq. Scale bar represents 500 $\mathrm{\upmu m}$. The leftmost column corresponds to sufficiently small Wi such that the flow is laminar. The righthand three columns show the unstable flow at three different time points. In (a--b) and (c--d), the flow is multistable, as in Fig.~\ref{300hpam}. By contrast, in (e--f), the flow is unstable, but is primarily ``eddy-free''. (g) Onset of unstable flow characterized again using the measure of eddy size fluctuations $\mathcal{A}\equiv A'_{\mathrm{eddy,rms}}/\langle A_{\mathrm{eddy}}\rangle_t$. (h) Onset of multistability characterized again using the measure of eddy size variations $\mathcal{E}\equiv\left(\mathrm{max}\{A_{\mathrm{eddy}}\}-\mathrm{min}\{A_{\mathrm{eddy}}\}\right)/A_{\mathrm{pore}}$. }
\end{figure*}

First, we test a higher concentration (900 ppm) of HPAM dissolved in the same Gl-Aq-DMSO solvent (crimson circle in Fig.~\ref{summary}). Unlike the dilute case of \S\ref{subsec:300_MS}, this solution is semi-dilute ($c/c^*\approx1.5$). The increased amount of polymer imparts further elasticity to the solution, and importantly, renders it moderately shear-thinning, as shown by the crimson point in Fig.~\ref{summary}. Notably, this added shear-thinning does \emph{not} abrogate multistability (Supplementary Movie 2). An example is shown in Fig.~\ref{streakline_3700_hq900_4500}(a--b): Pore body 16 [top row] is in the eddy-dominated unstable state, while pore body 17 [bottom row] is simultaneously in the eddy-free unstable state. Following \S\ref{subsec:300_MS}, we characterize this behavior by measuring the extent of unstable flow and multistability, $\mathcal{A}$ and $\mathcal{E}$, respectively, over a range of flow rates and for devices with varying pore constriction spacings. We find similar behavior to the cases described in \S\ref{subsec:300_MS}, as shown by the crimson points in Fig.~\ref{streakline_3700_hq900_4500}(g--h). For $l_s=W_0$, the elastic instability arises at $\dot{\gamma}_{w,c} \sim4-7  \: \mathrm{s^{-1}}$ ($\mathrm{Wi}\sim5-6$, $\mathrm{M}\sim19-31$) and multistability correspondingly arises above $\dot{\gamma}_{w,c} \sim7-34 \: \mathrm{s^{-1}}$ ($\mathrm{Wi}\sim6-10$, $\mathrm{M}\sim31-89$). Consistent with the idea that multistability arises when flow-induced polymer elongation is faster than the relaxation of polymers as they are advected between pore constrictions, this threshold is again shifted to larger shear rates with increasing $l_s$.

Next, we examine the generality of these findings---that shear-thinning does not abrogate multistability, which arises when polymers are stretched faster than they can relax between pore constrictions---by testing another highly shear-thinning and elastic semi-dilute polymer solution. In particular, we test the same Gl-Aq solution as the starting case of \S\ref{subsec:300_MS}, but at a higher concentration of 4500 ppm HPAM ($c/c^*\approx5$, green circle in Fig.~\ref{summary}), over a range of flow rates and pore constriction spacings. We again find similar behavior to the cases described in \S\ref{subsec:300_MS} and the case of 900 ppm HPAM, Gl-Aq-DMSO shown in Fig.~\ref{streakline_3700_hq900_4500}(a--b). Two exemplary pores are shown in Fig.~\ref{streakline_3700_hq900_4500}(c--d) and Supplementary Movie 1, and the aggregated measurements of $\mathcal{A}$ and $\mathcal{E}$ characterizing the extent of unstable flow and multistability are shown by the green points in Fig.~\ref{streakline_3700_hq900_4500}(g--h). As before, for sufficiently large shear rates, the flow becomes unstable and exhibits multistability. Moreover, this threshold is again shifted to larger shear rates as $l_s$ increases, further supporting the picture proposed in Refs.~\cite{browne_bistability_2020,kumar_numerical_2021}.

Finally, we test two shear-thinning but less elastic fluids. One is formulated by maintaining the same relative concentration $c/c^* \approx 5$ in the semidilute, unentangled regime, but with the polymer dissolved in ultrapure water, which acts as a higher-quality solvent (1000 ppm HPAM, Aq). The other is formulated using an even higher polymer concentration in the entangled regime ($c/c^* \approx 19$), again in ultrapure water (3700 ppm HPAM, Aq---orange star in Fig.~\ref{summary}). In this case, based on the picture proposed in Refs.~\cite{browne_bistability_2020,kumar_numerical_2021}, we expect that multistability will be suppressed because the characteristic solution relaxation times are reduced [Fig.~\ref{rheology}(e) and Table~\ref{tab:rheoparameters}]. Our experiments confirm this expectation (Supplementary Movie 2). Two exemplary pores are shown in Fig.~\ref{streakline_3700_hq900_4500}(e--f), and the aggregated measurements of $\mathcal{A}$ and $\mathcal{E}$ characterizing the extent of unstable flow and multistability are shown by the yellow and orange points in Fig.~\ref{streakline_3700_hq900_4500}(g--h). As before, for sufficiently large shear rates, the flow becomes unstable; however, it does not become multistable. Instead, all pores show the same behavior---the corner eddies become progressively smaller with increasing shear rate, and the flow exhibits strong spatiotemporal fluctuations throughout the pore body similar to the ``eddy-free'' case---even at the largest shear rates and smallest pore constriction spacings.

\subsection{\label{subsec:deborah}Streamwise Deborah number captures the onset of multistability}
\noindent Taken altogether, the experiments described in \S\ref{subsec:300_MS}--\ref{subsec:rheo_MS} demonstrate that shear-thinning does not abrogate the onset of the elastic instability and the resulting development of multistability. Moreover, we find no correlation between the onset of multistability and standard physicochemical descriptors of polymer solutions: the degree of shear-thinning $S_\textrm{max}$, relative concentration regime $c/c^*$, solvent quality, zero-shear viscosity $\eta_0$, or the solvent contribution to the total viscosity $\beta$. Instead, guided by the picture proposed in Refs.~\cite{browne_bistability_2020,kumar_numerical_2021}, we examine whether the streamwise Deborah number $\mathrm{De}=\lambda/t_{\mathrm{adv}}$ can capture the onset of multistability. We calculate the characteristic time for fluid to be advected between pore constrictions as $t_{\mathrm{adv}}=V_{\mathrm{pore}}/Q$. Importantly, unlike in our previous work~\cite{browne_bistability_2020}, here, the solution stress relaxation time $\lambda$ incorporates shear-thinning rheology through its rate dependence, as described in \S\ref{subsec:rheo}. Specifically, for the non-shear-thinning solutions, we take $\lambda=\lambda_0$, the longest stress relaxation time, as in previous work ~\cite{boger_viscoelastic_1987,james_boger_2009}; this choice quantifies the expectation that for these solutions, the longest relaxation time corresponding to relaxation of an entire polymer chain governs stress relaxation~\cite{ bird_dynamics_1987, rubinstein_polymer_2003,morozov_introductory_2007,larson_modeling_2015}. By contrast, for the shear-thinning fluids, we evaluate the relaxation time as $\lambda=\lambda_{\mathrm{WM}}(\dot{\gamma}_{w,c})$ at each flow rate tested; this choice quantifies the expectation that these solutions have multiple modes of stress relaxation that are coupled to flow-induced microstructural rearrangements~\cite{de_gennes_reptation_1971,rubinstein_polymer_2003,litvinov_mesoscopic_2011,klebinger_transient_2013,howe_flow_2015,larson_modeling_2015}.

All of our results---across eight different test fluids of systematically-varying rheological properties and three different pore constriction spacings---are summarized by the state diagram shown in Fig.~\ref{1D_state}. Remarkably, despite the complex nature of the elastic instability, all of our results show excellent collapse when parameterized by $\mathrm{De}$. When $\mathrm{De}<1$, we do not observe multistability (open symbols): the flow is either laminar, or with all pores exhibiting similar unstable ``eddy-free'' flow. By contrast, when $\mathrm{De}$ exceeds a threshold value $\mathcal{O}(1)$, the flow is multistable (closed symbols). This collapse therefore demonstrates that the picture proposed by Refs.~\cite{browne_bistability_2020,kumar_numerical_2021}---that multistability arises when flow-induced polymer elongation is faster than polymer relaxation between adjacent pore constrictions---holds across a diverse array of fluids with varying rheological properties.

\section{Conclusion}
\noindent In summary, using flow visualization in microfabricated millifluidic devices, we have investigated the influence of systematic variations in fluid shear-thinning or elasticity on the unstable flow of polymer solutions in 1D ordered arrays of pore constrictions. In all cases, when the fluid is sufficiently elastic, it exhibits a flow instability above a threshold flow rate; intriguingly, the parameters $\mathrm{Wi}$ and $\mathrm{M}$ do not appear to uniquely capture the onset of this unstable flow across the different solutions and device geometries. However, the parameter $\mathrm{De}$ --- which compares the  shear rate-dependent longest stress relaxation time to the advection time between pores --- captures the onset of multistability in the unstable flow across all our experiments. Our work thereby demonstrates that the picture proposed in Refs.~\cite{browne_bistability_2020,kumar_numerical_2021} holds more broadly, and corroborates other studies suggesting that the rate-dependence of polymer relaxation can influence elastic instabilities~\cite{casanellas_stabilizing_2016,shakeri_scaling_2022}. 

Our experiments explored polymer solutions of varying concentrations, solvent qualities, and viscosities, primarily using the same high molecular weight HPAM. In future work, it will be useful to investigate variations in the polymer molecular weight, architecture, and composition to further explore the general applicability of our findings. It will also be useful to examine different definitions of the streamwise Deborah number --- for example, using shear stress relaxation times determined using other approaches than that used here~\cite{del_giudice_relaxation_2017,soulies_characterisation_2017,khalkhal_analyzing_2022,khalkhal_analyzing_2022} or, given the central role of polymer extension in determining the characteristics of the unstable flow behavior, using an extensional relaxation time~\cite{klebinger_transient_2013,larson_modeling_2015} determined using e.g., capillary breakup rheometry~\cite{mckinley_visco-elasto-capillary_2005}, cross slot rheometry~\cite{haward_optimized_2012,haward_extensional_2023,haward_extensional_2023-1}, or dripping-on-substrate rheometry~\cite{dinic_pinch-off_2017}.

Altogether, by deepening understanding of the influence of fluid rheology on elastic instabilities, our work helps to pave the way towards the rational tuning of both fluid rheology~\cite{ewoldt_designing_2022} and porous medium geometry~\cite{stone_engineering_2004} to harness such instabilities in diverse chemical, energy, environmental, and industrial settings---for example, using them to enhance heat/mass transport in porous media~\cite{groisman_efficient_2001,traore_efficient_2015,whalley_enhancing_2015, browne_homogenizing_2023,kumar_stress_2023,browne_harnessing_2023}, where eddy-dominated pores could act as semi-compartmentalized microreactors~\cite{trojanowicz_flow_2020}.

\section*{SI Movie Captions}
\noindent\textbf{SI Movie 1.} Comparison of laminar and unstable flow behavior in the $l_s = 1W_0$ 1-D pore array. (a-b) Streakline imaging of the flow of the weakly shear-thinning, highly elastic 300 ppm HPAM, Gl-Aq solution in pore body 16 showing (a) laminar (Wi = 1.8 $<$ Wi$_c \approx 6$) and (b) unstable (Wi = 26 $>$ Wi$_c \approx 6$) flow behavior. Flow is from left to right. (c-d) Streakline imaging of the flow of the highly shear-thinning, highly elastic 4500 ppm HPAM, Gl-Aq solution showing (c) laminar (Wi = 10 $<$ Wi$_c \approx 15$) and (d) unstable (Wi = 34 $>$ Wi$_c \approx 15$) flow behavior. Unstable flow behavior is visualized by crossing streaklines and fluctuations in the eddy sizes. Both solutions exhibit multistability in the unstable flow state, as described in the main text; thus, shear-thinning does not abrogate multistability. The movies shown correspond to an ``eddy-dominated'' pore over the imaging window. Streaklines are generated using a moving average of the fluorescence intensity of tracer particles over 30 successive frames. Scale bar is $500 \: \upmu$m. Videos play at 15x real time.\\

\noindent\textbf{SI Movie 2.} Comparison of multistable and non-multistable flow behavior above the onset of elastic instability (Wi $>$ Wi$_c$) for polymer solutions in the $l_s = 1W_0$ 1-D pore array. (a-c) Unstable flow exhibiting multistability for the moderately shear-thinning, highly elastic 900 ppm HPAM, Gl-Aq-DMSO solution in three consecutive pore bodies. Pore bodies 15 and 16 appear eddy-dominated over the imaging window, while pore body 17 transitions between eddy-dominated and eddy-free states. Flow is from left to right. (d-f) Unstable flow showing no multistability for the highly shear-thinning, moderately elastic 3700 ppm HPAM, Aq solution. The unstable flow is observed qualitatively by crossing streaklines in the bulk of the pore body and fluctuations of the small corner eddies. However, all pore bodies exhibit the same instability flow behavior at the same imposed flow rate, showing a lack of multistability. Imaging is recorded sequentially in adjacent pores, producing a time offset of 2 minutes from pore-to-pore. Streaklines are generated using a moving average of the fluorescence intensity of tracer particles over 30 successive frames. Scale bar is $500 \: \upmu$m. Videos play at 15x real time.

%%%%%%%%%%%%%%%%%%%%%%%%%%%%%%%%%%%%%%%%%%%%%%%%%%%%%%%
\begin{acknowledgments}
\noindent It is a pleasure to acknowledge insightful discussions with Anna Hancock and Christopher Browne, and the use of Princeton's Imaging and Analysis Center (IAC), which is partially supported by the Princeton Center for Complex Materials (PCCM), a National Science Foundation (NSF) Materials Research Science and Engineering Center (MRSEC; DMR-2011750). We also acknowledge funding support from the Camille Dreyfus Teacher-Scholar Program.
\end{acknowledgments}

\section*{Author Contributions}
\noindent E.Y.C. and S.S.D. designed the experiments; E.Y.C. performed all experiments; E.Y.C. and S.S.D. analyzed all data, discussed the results and implications, and wrote the manuscript; and S.S.D. designed and supervised the overall project.

\section*{Conflict of Interest Statement}
\noindent There are no conflicts of interest to declare.

\section*{Data Availability Statement}
\noindent All data, and description of all methods required to reproduce the results, are completely included in the manuscript and/or supporting information.

\nocite{*}

\end{document}